\documentclass{elsart}

\usepackage{epsfig}

\usepackage{amssymb}

\begin{document}
\begin{frontmatter}
\title{Tracking Performance of the Scintillating Fiber Detector
in the K2K Experiment}

\author[SNU]{\mbox{B. J. Kim}\corauthref{Contact}},
\author[Kobe]{\mbox{T. Iwashita}},
\author[KEK]{\mbox{T. Ishida}},
\author[KEK]{\mbox{E. J. Jeon}},
\author[SUT]{\mbox{H. Yokoyama}},
\author[Kobe]{\mbox{S. Aoki}},
\author[UW]{\mbox{H. G. Berns}},
\author[SNU]{\mbox{H. C. Bhang}}, 
\author[UW]{\mbox{S. Boyd}},
\author[Kobe]{\mbox{K. Fujii}},
\author[Kobe]{\mbox{T. Hara}},
\author[KEK]{\mbox{Y. Hayato}},
\author[SUNY]{\mbox{J. Hill}},
\author[KEK]{\mbox{T. Ishii}},
\author[KEK]{\mbox{H. Ishino}},
\author[SUNY]{\mbox{C. K. Jung}},
\author[BU]{\mbox{E. Kearns}},
\author[SNU]{\mbox{H. I. Kim}},
\author[SNU]{\mbox{J. H. Kim}},
\author[CNU]{\mbox{J. Y. Kim}},
\author[SNU]{\mbox{S. B. Kim}},
\author[KEK]{\mbox{T. Kobayashi}},
\author[Hawaii]{\mbox{S. Matsuno}},
\author[UCI]{\mbox{S. Mine}},
\author[KEK]{\mbox{K. Nakamura}},
\author[Niigata]{\mbox{M. Nakamura}},
\author[Kyoto]{\mbox{K. Nishikawa}},
\author[Kobe]{\mbox{T. Otaki}},
\author[KEK]{\mbox{Y. Oyama}},
\author[KEK]{\mbox{H. Park}\thanksref{Parknow}},
\author[KEK]{\mbox{M. Sakuda}},
\author[BU]{\mbox{K. Scholberg}},
\author[SUNY]{\mbox{E. Sharkey}},
\author[BU]{\mbox{J. L. Stone}},
\author[Kobe]{\mbox{A. Suzuki}},
\author[Kobe]{\mbox{K. Takenaka}},
\author[Niigata]{\mbox{N. Tamura}},
\author[Kobe]{\mbox{Y. Tanaka}},
\author[BU]{\mbox{C. W. Walter}},
\author[UW]{\mbox{J. Wilkes}},
\author[SNU]{\mbox{J. Yoo}}, and
\author[Osaka]{\mbox{M. Yoshida}}
\address[BU]{Department of Physics, Boston University, Boston, MA 02215, USA}
\address[CNU]{Department of Physics, Chonnam National University, Kwangju 500-757, Korea}
\address[KEK]{High Energy Accelerator Research Organization, Tsukuba, Ibaraki 305-0801, Japan}
\address[UCI]{Department of Physics and Astronomy, University of California, Irvine, CA 92697-4575, USA}
\address[Hawaii]{Department of Physics and Astronomy, University of Hawaii, Honolulu, HI 96822, USA}
\address[Kobe]{Kobe University, Kobe, Hyogo 657-8501, Japan}
\address[Kyoto]{Department of Physics, Kyoto University, Kyoto 606-8502, Japan}
\address[Niigata]{Department of Physics, Niigata University, Niigata, Niigata 950-2181, Japan}
\address[SUNY]{Department of Physics and Astronomy, State University of New York, Stony Brook, NY 11794-3800, USA}
\address[Okayama]{Department of Physics, Okayama University, Okayama, Okayama 700-8530, Japan}
\address[Osaka]{Department of Physics, Osaka University, Toyonaka, Osaka 560-0043, Japan}
\address[SUT]{Department of Physics, Science University of Tokyo, Noda, Chiba 278-0022, Japan}
\address[SNU]{Department of Physics, Seoul National University, Seoul 151-742, Korea}
\address[UW]{Department of Physics, University of Washington, Seatle, WA 98195-1560, USA}
\corauth[Contact]{Corresponding author. Tel.: +82-2-872-6868; fax: +82-2-884-3002.\\ E-mail address: bockjoo@neutrino.kek.jp (B. J. Kim).}
\thanks[Parknow]{Present address: Department of Physics, Seoul National University, Seoul 151-742, KOREA}
\begin{abstract}
\baselineskip 15pt   

The K2K long-baseline neutrino oscillation experiment uses 
a Scintillating Fiber Detector (SciFi) to
reconstruct charged particles produced in neutrino
interactions in the near detector. We describe the
track reconstruction algorithm and the performance of the
SciFi after three years of operation.
\end{abstract}       


\end{frontmatter}

\section{K2K Experiment}

The K2K experiment is the first 
long-baseline neutrino oscillation experiment in operation.
Muon neutrinos($\nu_\mu$) are
produced via the decay of secondary pions from interactions of the 
KEK 12-GeV proton beam in an aluminum target. The neutrino beam is
directed through a near detector system at KEK to the Super-Kamiokande
underground detector\cite{SK} 250km away.
By comparing the neutrino fluxes and energy spectra measured at the near
detector and at Super-Kamiokande, we are able to investigate neutrino 
oscillations. Details of the K2K experiment have been presented elsewhere
\cite{detect00}.

\begin{figure}[h]
\begin{center}
\vspace{-2.0cm}
\mbox{\epsfig{figure=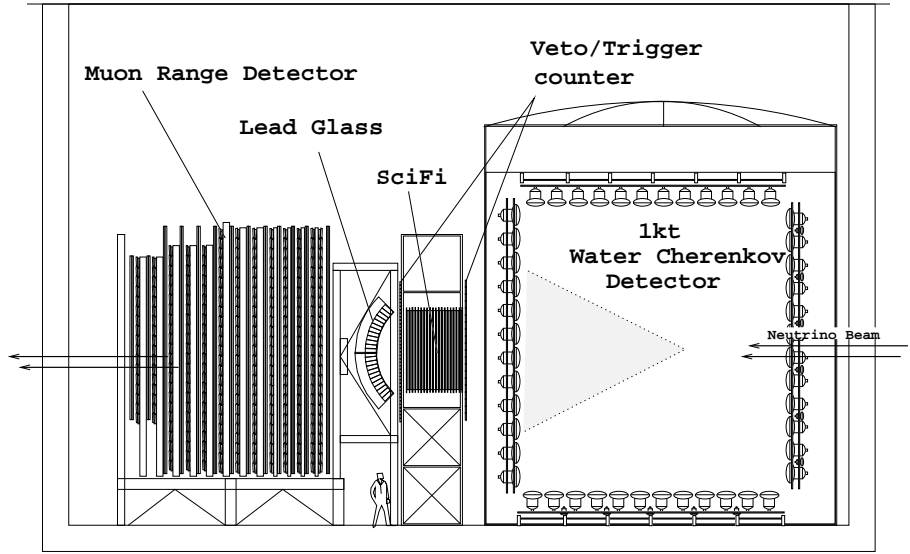,width=12cm}}
\vspace{-2.0cm}
\caption{The layout of the K2K near detector. It consists of
a 1-kt water Cherenkov detector, a scintillating fiber detector,
a hodoscope of plastic scintillation counters, a lead glass detector,
and a muon range detector.}
\label{k2knear}
\end{center}
\end{figure}

The near detector, located 300 m downstream of the pion production point,
consists of a 1kt water Cherenkov detector, a scintillating fiber 
tracking detector (SciFi),
a hodoscope of plastic scintillation counters (VETO) around the SciFi, 
a lead glass detector (LG), 
and a muon range detector (MRD)~\cite{mrd}. It is illustrated in Fig.~\ref{k2knear}. 
%
%
The main purpose of the near detector is to measure the 
profile, flux, and energy spectrum of the neutrino beam.
Among the near detector components, the SciFi 
can provide tracking capability good enough
to discriminate neutrino interaction types and 
to measure energy and cross sections of the muon neutrino.

The K2K experiment started taking data in April, 1999. 
This paper will describe the performance of the Scintillating Fiber Detector 
after three years of  operation. 
A description of the K2K scintillating fiber detector is given in 
Section~\ref{ovrscifi}.
Hit reconstruction algorithm and performance are described in Section~\ref{hitrec}.
Track and vertex reconstruction is described in Section~\ref{trackrec}.
Performance of the neutrino event reconstruction algorithms are presented in 
Section~\ref{neutrec}.
We conclude in Section~\ref{concls}.

\section{Scintillating Fiber Detector}
\label{ovrscifi}
%
%
\begin{figure}[ht]
\begin{center}
\vbox{{
\psfig{figure=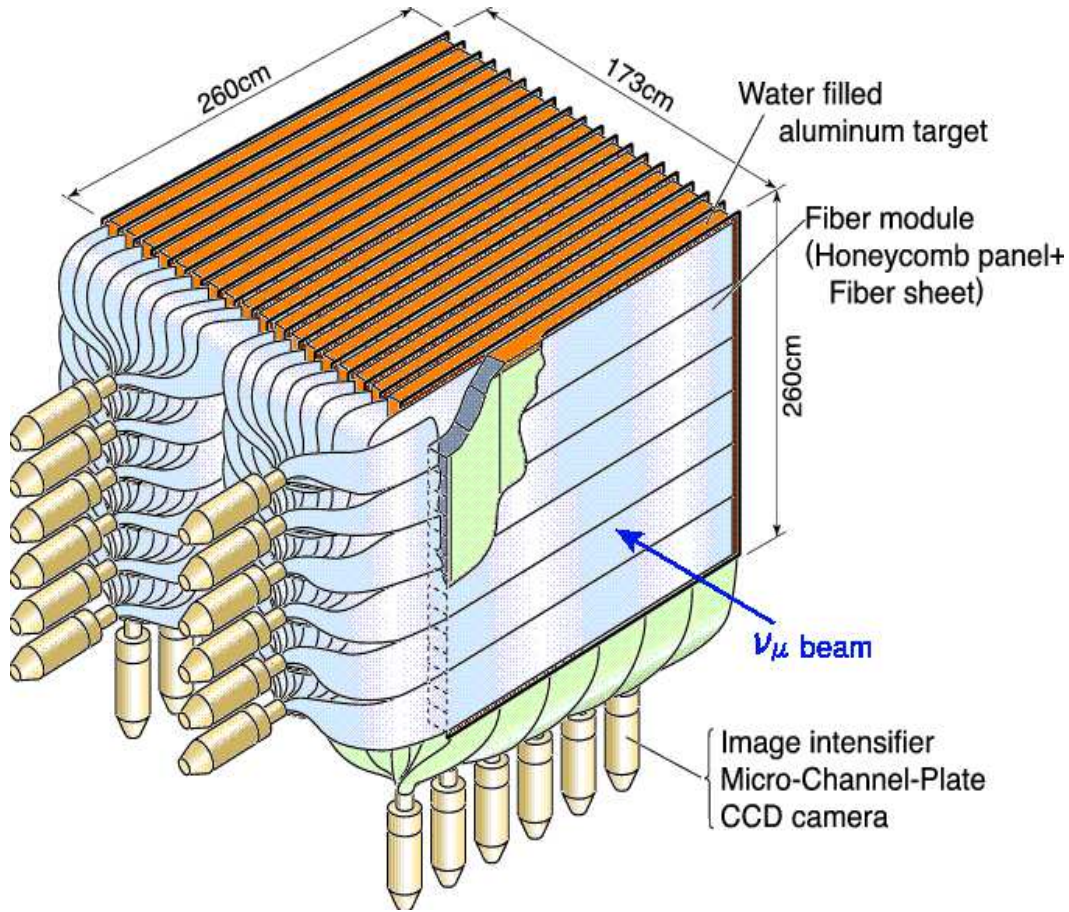,width=14cm,height=12cm}
      }}
\caption{A schematic view of the K2K scintillating fiber 
detector. The detector has the dimension of 
260cm$\times$260cm$\times$173cm in ($x,y,z$). 
The central fiducial mass region is defined 
to be 220cm$\times$220cm$\times$159cm in ($x,y,z$).}
\label{k2kscifi}
\end{center}
\end{figure}
The SciFi is a charged particle tracking detector using scintillating
fiber technology. Scintillating fiber detectors have been used 
for charged particle trackers
by other experiments~\cite{UA2CHORUSH1}. 
A scintillating fiber detector was chosen for this application since
it was necessary to have high tracking efficiency for reconstructing 
charged particles and their vertex positions when more than two visible particles 
are produced in neutrino interactions.
Using the detector, we can separate quasi-elastic events from 
inelastic events, and define the fiducial volume of the neutrino interactions
with 1\% accuracy. 

The water is contained in a thin
aluminium structure, and the water layers are interleaved with
scintillating fiber layers. Water is chosen as a target material 
in order to minimize systematic uncertainties in neutrino cross-section
when the event rates are compared between the SciFi and Super-Kamiokande detectors.
A scintillating fiber layer provides the hit-position 
measurement in both horizontal and vertical 
directions~\footnote{In the K2K coordinate system, the $z$-axis is along 
the nominal neutrino beam direction. 
$y$-axis is the vertical direction pointing
upwards, and $x$-axis is the horizontal direction.
}.
An individual scintillation sheet has a thickness of 1.3mm, consisting of
two staggered layers of scintillating fibers. 
Each fiber has a diameter of 0.7mm.
The vertical and horizonal fiber sheets are separated by a 1.6-cm thick honeycomb panel.
There are 20 scintillating fiber layers in total, and 
the adjacent layers are spaced 9cm apart along the neutrino beam direction. 

One layer of 240cm$\times$240cm$\times$6cm water target, consisting of 
15 aluminium tanks, is interleaved with two fiber layers. 
The wall thickness of the aluminium tank is 1.8mm.
The weights of water, aluminium, and the honeycomb panel are 
6.0tons, 1.4tons, and 0.8tons, respectively.
A schematic view of the SciFi is shown in Fig.~\ref{k2kscifi}.
 
For processing light signals, 
a group of fibers are bundled and 
coupled to an image intensifier tube (IIT), 
an optical lens, and a CCD camera as shown in Fig.~\ref{k2kscifi}.
A total of 24 IIT's are used to read out all 274080 fibers in the SciFi.  
In order to find the correspondence between the hit fiber and the
CCD images, 
a certain number of selected fibers are illuminated periodically at the opposite side of IIT's by an electro-luminescent (EL) plate.
This is referred to as the EL calibration. A more detailed description of 
the SciFi detector is found elsewhere~\cite{K2KSciFiNIM1}. 

\section{Hit reconstruction}
\label{hitrec}
\subsection{Hit Finding Algorithm}
\label{hitrecon}
%
%
\begin{figure}[tbp]
  \begin{center}
    \epsfig{file=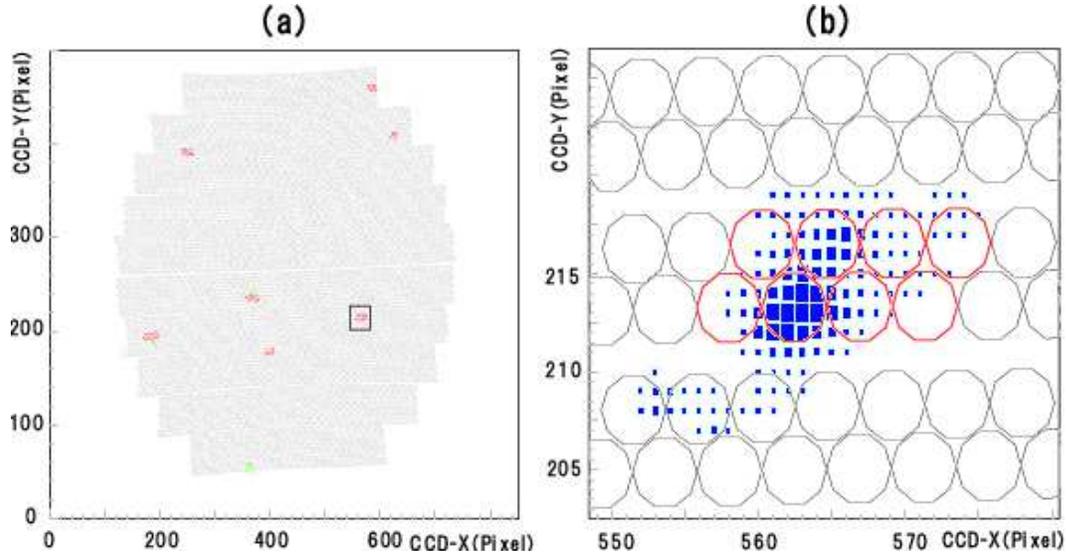,width=14.cm}
    \caption{(a) A typical CCD image of an IIT from a cosmic-ray event.
     Small distortions of the CCD image are introduced due to the position
     dependence of pixel size at the IIT-CCD contact.
     (b) An expanded view of (a) for a typical SciFi hit.
     The size of hit pixel is proportional to brightness in the ADC. 
     Hit fibers are represented by the thick circles.
     A ``SciFi hit'' is defined as a cluster of neighboring hit fibers.  
     }
    \label{fig:hitfind}
  \end{center}
\end{figure}
The CCD pixel images need to be handled to reconstruct hits in the SciFi.
Fig.~\ref{fig:hitfind}(a) shows a typical CCD image of an IIT from 
a cosmic-ray event.
A fiber bundle, consisting of ten fiber sheets, is arranged to cover a sensitive 
region of an IIT effectively.
Fig.~\ref{fig:hitfind}(b) is an expanded view of (a), 
illustrating a typical ``SciFi hit'', when a cosmic-ray muon 
passes through the fiber sheet.
The hit reconstruction is performed in the following four steps:
\begin{description}
\item[1. Loading Hit Pixels~~] 
   The raw data of the SciFi consist of hit pixels in the CCD coordinates ($x$, $y$)  
   and 8 bits of ADC brightness. These are all encoded in one byte per pixel.
   Typically, there are 3,500 hit pixels in a neutrino event.
\item[2. Finding Pixel Clusters~~] 
   A pixel cluster is determined by a group of neighboring hit pixels. 
   At this stage, isolated single pixel hits, coming from random electrical
   noise of CCD, are rejected.
\item[3. Identifying Hit Fibers~~] 
   A hit fiber is identified by requiring at least one or two 
   hit pixels of cluster within the fiber. 
   The hit fiber position in the CCD coordinate is determined from a mapping
   table that is obtained from the EL calibration.
\item[4. Making Hit Fiber Clusters~~] 
   A fiber cluster is determined by a group of neighboring hit fibers.
\end{description}
We apply two kinds of filtering algorithms to the fiber clusters to improve 
the signal-to-noise ratio:
\begin{description}
\item[$\bullet$~~]  We require that a fiber cluster should consist of
    hit fibers in both upper and lower layers of a fiber sheet. 
    This eliminates fiber clusters of
    small pixel images which come from dark current in an IIT.  
\item[$\bullet$~~]  When a big pixel cluster lies on the adjacent
    fiber sheets in a bundle and makes multiple fiber 
    clusters, we choose 
    only fiber clusters containing peak of ADC brightness.
    This removes fake fiber-clusters introduced
    by the dense arrangement of fiber sheets in a bundle.
\end{description}
We consider a fiber cluster passing through these two filters 
as a SciFi ``hit''. Every hit position in the CCD coordinate is then 
converted into the position in the detector 
coordinates ($x,z$ or $y,z$). The hit position is defined as the centroid of
a fiber cluster, weighted by the ADC value of the hit pixel.

\subsection{Number of pixels per hit($N_{pix}$) and its corrections} 

%
%
\begin{figure}[tbp]
  \begin{center}
    \epsfig{file=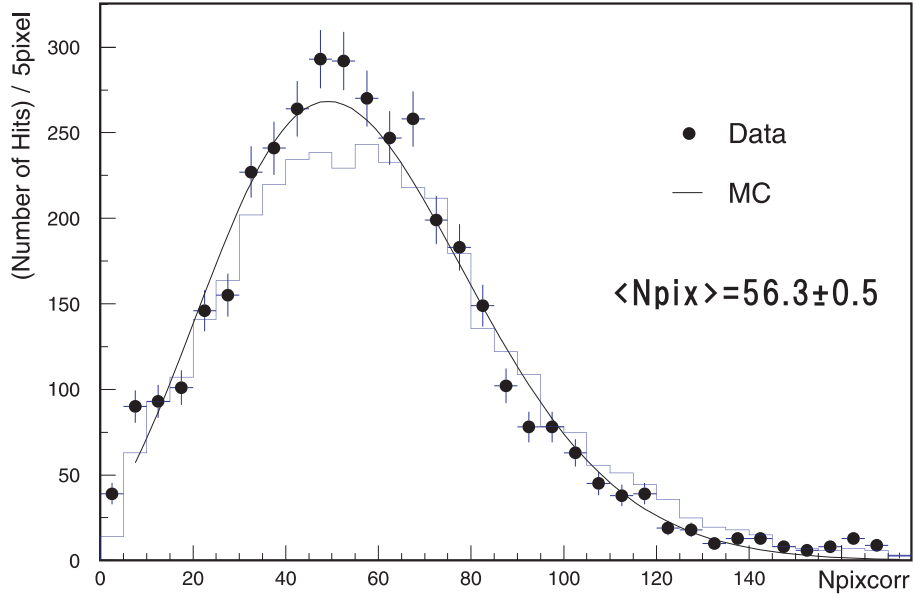,width=12cm}
    \caption{An $N_{pix}$ distribution
             with the corrections applied for both attenuation and dependence on the 
             incident angle
             as given in Eq.(\ref{eq:npixcorr}).
     }
    \label{fig:ii10npix}
  \end{center}
\end{figure}
A found hit gives the
number of pixels for the parent pixel cluster, $N_{pix}$.
LED calibration data show that $N_{pix}$ has good linearity 
with respect to the number of photons injected to IIT. 
Since the intensity of observed light
is proportional to the energy loss in the fiber, $N_{pix}$ is expected
to be a good estimator of the energy of the parent particle. 

Fig.~\ref{fig:ii10npix} shows an $N_{pix}$ distribution 
for cosmic-ray muons. Based on a reconstructed track, $N_{pix}$ can be corrected
to account for both the attenuation of light in the fiber, and the
dependence of light yields on the incident angle ($\theta_{in}$) of a particle
on a fiber sheet:
\begin{eqnarray}
N_{pixcorr}
 = \frac{F(l_0)}{F(l)} \times cos\theta_{in} \times N_{pix}.
\label{eq:npixcorr}
\end{eqnarray}
The first term is the correction for attenuation of the scintillation light 
in the fiber and reflection at the far end of fiber 
coated with aluminum~\cite{K2KSciFiNIM1}. The correction is normalized by the 
value in case of the light from the detector center ($l_0$=130$cm$). 
Thus, $F(l)$ is given by:
\begin{eqnarray}
F(l)= exp(-\frac{l+l_g}{\lambda}) 
           + R\cdot exp(-\frac{2\cdot L-l+l_g}{\lambda}),
\label{eq:travel}
\end{eqnarray}
where $l$ is the distance from the hit position to the near end of a fiber,
$l_g($=100$cm$) is light guide length from the near end to the IIT, and
$\lambda($=323$cm$) is the attenuation length.
$R$(=0.74) is the reflection coefficient at
the far end of a fiber, and $L$(=254$cm$) is the length of a fiber. 

The second term in Eq.~(\ref{eq:npixcorr}) is the correction 
for the dependence of light yields on the incident angle 
with respect to the fiber sheet. Here, $\theta_{in}$=0 is perpendicular 
to the fiber sheet.
$N_{pixcorr}$ is thus equivalent to $N_{pix}$ when a particle
passes fiber sheet at the center of the detector $x/y$ planes with 
$\theta_{in}$=0.

\begin{figure}[tbp]
  \begin{center}
    \epsfig{file=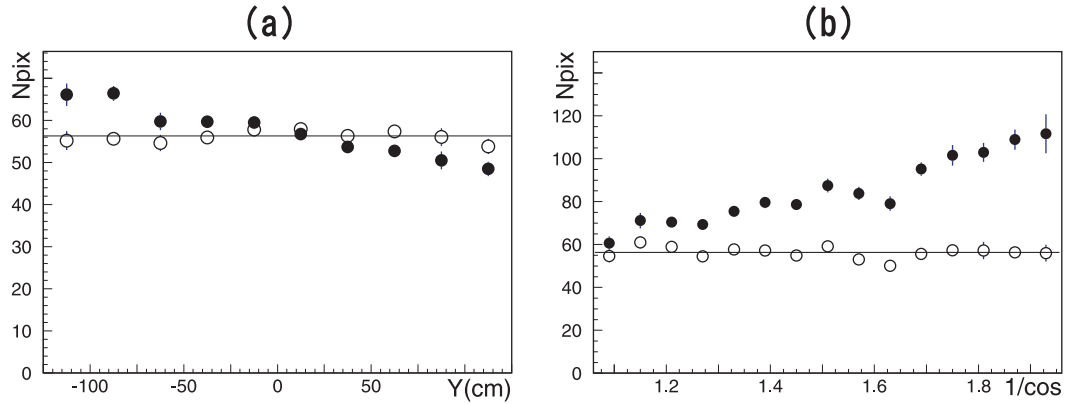,width=14cm}
    \caption{An average value of $N_{pix}$, $\langle N_{pix} \rangle$,  
             as a function of
             (a) hit position along the fiber and
             (b) 1/$cos\theta_{in}$, where $\theta_{in}$ is the incident 
             angle with respect to the fiber sheet, respectively. 
             Solid (open) circles are before (after) the correction.
             The negative $x,y$ coordinates in (a) correspond to positions
             close to readout end of the fiber, and a
             larger value of 
             $\langle N_{pix} \rangle$ is expected before the correction.        
             }
    \label{fig:npixcorr}
  \end{center}
\end{figure}

Figs.~\ref{fig:npixcorr}(a) and (b) show average values of $N_{pix}$  
($\langle N_{pix} \rangle$) 
\footnote{
We employ the following Poisson function to fit the $N_{pix}$ distribution:
$$F(N_{pix}; A, \alpha, \langle N_{pix} \rangle)= A \cdot 
  \frac{e^{-\mu} \times \mu^x }
       {\Gamma(x+1)},\ \ \mu \equiv \langle N_{pix} \rangle/\alpha,\ \ 
   {\rm and}\ \ x \equiv N_{pix}/\alpha, $$
where $A$ is an overall normalization factor and $\alpha$ is a 
conversion factor.  $\langle N_{pix} \rangle$ is the
average value of the $N_{pix}$ distribution.
}
as a function of hit position along the fiber  
and as a function of $1/cos(\theta_{in})$, respectively.
%

\subsection{Simulation of CCD hit pixels}

To reproduce SciFi hits accurately in the K2K Monte Carlo simulation,
we include
simulation of hit pixels on CCD due to charged 
particles and noise in the following three steps:
\begin{description}
\item[1.~]
Charged particles are tracked through the detector
using GEANT\cite{GEANT}. The energy loss of a particle traversing
a fiber is computed and then converted to the number of photoelectrons
observed by the IIT/CCD imaging system, taking into account attentuation
along the fiber, reflection by the aluminium coated at the end of a fiber, 
and the quantum efficiency of the IIT photocathodes.
\item[2.~]
The photoelectrons are converted to an image on the CCD camera surface.
Individual photoelectrons are distributed around the fiber center according to
a gaussian distribution with parameters determined using data of a single 
photoelectrons from LED calibration run. The size of the pixel cluster 
corresponding to the single photoelectron is then chosen from distributions of 
the LED data, and the pixel cluster is simulated accordingly.
The energy scale 
is tuned for each IIT using cosmic-ray muon data.
\item[3.~]
Random noise in a pixel cluster is generated for each IIT based 
on neutrino-beam data.
\end{description}

Two kinds of noise pixels are generated.
One is isolated single hit pixels
around the main pixel cluster of a charged track
in step {\bf 2.}.
The other is random noise pixels in a pixel cluster generated in step {\bf 3.}. 
The total number of noise hits averaged over IITs
is 1.67 for MC and 1.72 for cosmic-ray data. 
Based on the pixel simulation, the average number of noise hits is expected
to be 70 per event.
In the cosmic-ray data we find 
52 hits per event due to the random noise of CCD, and 17 hits 
per event due to the beam associated background.
The noise rate is quite low, considering the total number of fibers (274,080).

\subsection{Efficiency and Stability of Hit Reconstruction}

%
%
\begin{figure}[tbp]
  \begin{center}
    \epsfig{file=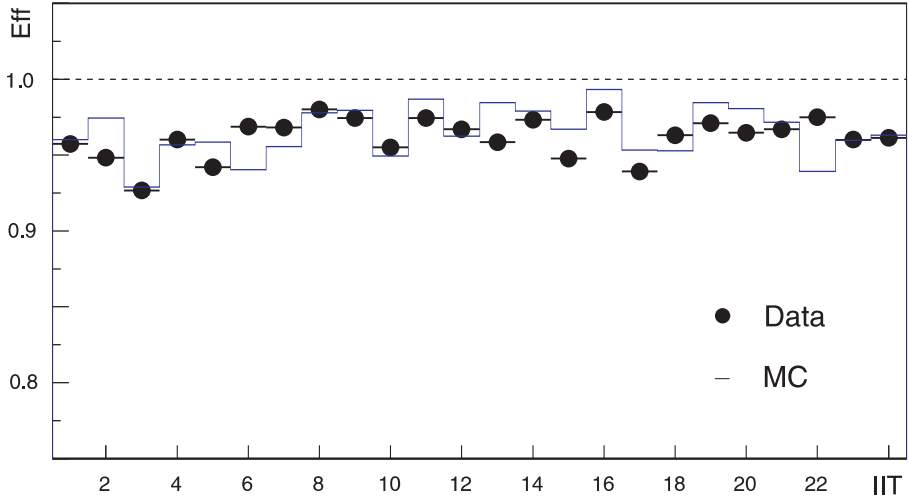,width=12cm}
    \caption{Hit finding efficiency $\epsilon_{hit}$ obtained from 
     cosmic-ray data}
    \label{fig:eff}
  \end{center}
\end{figure}

We can obtain the hit finding efficiency
$\epsilon_{hit}$ which is defined as:
\begin{eqnarray}
\epsilon_{hit} = \frac{\rm Number\ of\ observed\ hits}
                      {\rm Number\ of\ expected\ hits}, 
\end{eqnarray}
where the expected number of hits (denominator) is obtained from 
the information of reconstructed tracks.
Fig.~\ref{fig:eff} shows $\epsilon_{hit}$ for each IIT.
Data and Monte Carlo shows a good agreement.

Cosmic-ray data are taken during the normal data-taking period, roughly
once every two weeks. 
The efficiency is monitored 
using the data as a check of detector stability. 

We have performed a similar check on the stability of hit finding efficiency 
using the neutrino data. 
Clean single track events are selected by
requiring a track with the number of hit layers larger than nine, and its matching to
activity in the downstream VETO counter.

%
%
\begin{figure}[tbp]
  \begin{center}
    \epsfig{file=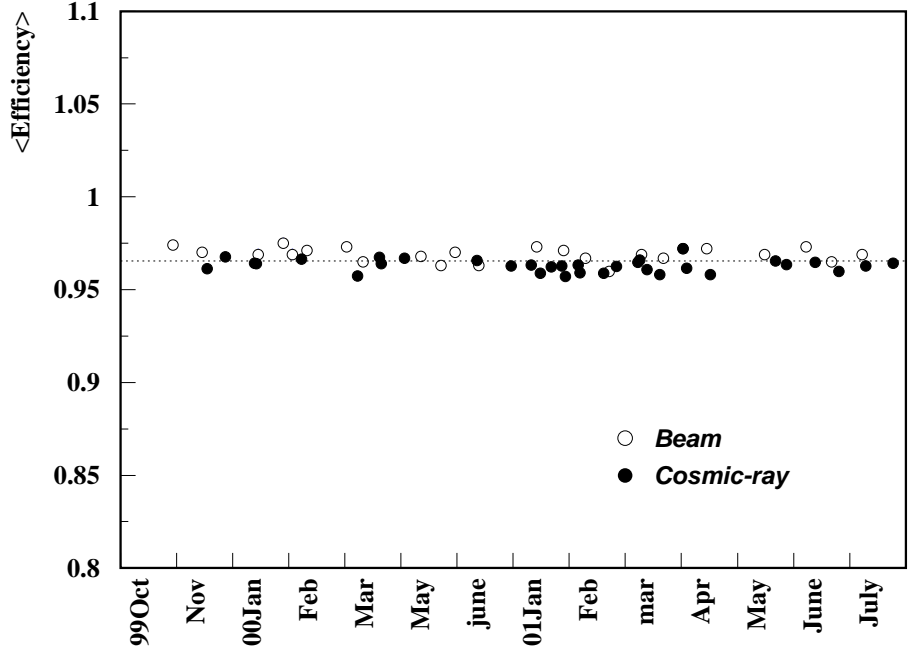,width=12cm}
    \vspace{0.2in}
    \caption{
    Hit finding efficiency as a function of data taking period for the
    neutrino data (open circle) and cosmic ray data (solid circel). 
    The dotted line represents an MC predicted hit finding efficiency.}
    \label{fig:stability}
  \end{center}
\end{figure}
Fig.~\ref{fig:stability}
shows hit finding efficiency, which is obtained by taking average
over all IIT's,
as a function of data-taking period.
The dotted line
at $96.8\%$ represents an MC prediction.
For the cosmic-ray data, we find an average value of $\epsilon_{hit}$ to be
$96.5\pm0.5\%$ for all IIT's and all time periods.
\footnote{ 
This is in good agreement with a simple estimation. When
a minimum ionizing particle crossing the fiber sheet creates
$\sim 8 p.e.$ (4$p.e.$ each for upper and lower layers),
hit coincidence probability between upper and lower layers
is $(1.-e^{-4})^2$=0.96. The requirement of the
coincidence between two fiber layers is chosen to 
keep the noise hits at the level of 70 hits per event.
}.

\section{Reconstruction of Track and Vertex}
\label{trackrec}
\subsection{Track Reconstruction Algorithm}
\label{trekfinding}
The track finding algorithm in the SciFi is optimized for the
charged current interactions which contain a muon in the final state.
The track finding algorithm consists of three steps.

First, the track finder searches for 2-dimensional (2D) track candidates in
the $(x,z)$ or $(y,z)$ planes which traverse at least three layers of SciFi. 
A 3-dimensional track is reconstructed by selecting the best combination 
of two 2D tracks.
In order to reduce the combinatorial fake tracks, the starting and 
ending hit layers are required not to be different by more than one layer, and
the overlapping of the two 2D tracks is required to be at least two layers.

Secondly, the combinatorial background or noise tracks are removed using other
components of the near detector (VETO and MRD). A track candidate in the
SciFi is required to match with both a hit in the VETO and a track in the MRD.

Once a good primary track is found,
the most upstream hit-position
of the primary track is
considered as a neutrino interaction point. 
The track finder starts
searching for extra tracks near the neutrino interaction point.
These tracks are mostly short tracks
with two or more SciFi hits. 
This method of extra track finding turns out to be an effective way 
to find short tracks of protons and/or pions from the neutrino
interactions. All the reconstructed tracks are subjected to 
a ``global fit'', a straight line fit to the found hits.

\subsection{Alignment and Position Resolution}

The geometry of the SciFi and its support structure was  
surveyed by mechanical means after their installation was completed. 
Based on the mechanical survey,
the alignment of fibers is estimated to 
be good to within 5mm. 

A better alignment was obtained
by minimizing the sum of the squared 
hit-residuals using the global fit for
cosmic ray tracks.
From the mean shift of the hit residuals,
we find the accuracy of the alignment to be $\sim$400$\mu$m. 

Multiple scatterings occurring in the SciFi
is not negligible compared with the size of a fiber. 
A Kalman filtering technique~\cite{kalman}~\cite{hepkal}~\cite{fruwirth} can be employed
to take the multiple scattering into account and, in principle, 
removes the multiple scattering effect during track fitting.
The technique calculates the error of a track fit relatively quickly,
and avoids the tedious matrix inversion by the global fit.
A track fit using the Kalman filtering technique is used 
to estimate the detector
resolution more accurately and to obtain a better alignment.
The implementation adopts the prescriptions described in 
Refs.~\cite{fruwirth} and ~\cite{kalappl}. 

Some curvature was introduced in the fiber sheets due to bending when they were
glued to a fiber module. 
Production of the fiber modules is described in Ref.~\cite{K2KSciFiNIM1}.
The curvature of a fiber sheet causes displacement from the aligned fiber-position
and results in a rotational misalignment.
The bending of every fiber sheet was measured after each fiber module was assembled.
The curvature of a fiber sheet was then obtained by fitting the measured fiber bendings to
a forth-order polynomial, and used to correct the rotational misalignment.

\begin{figure}[h]
\begin{center}
\vbox{
      {\psfig{figure=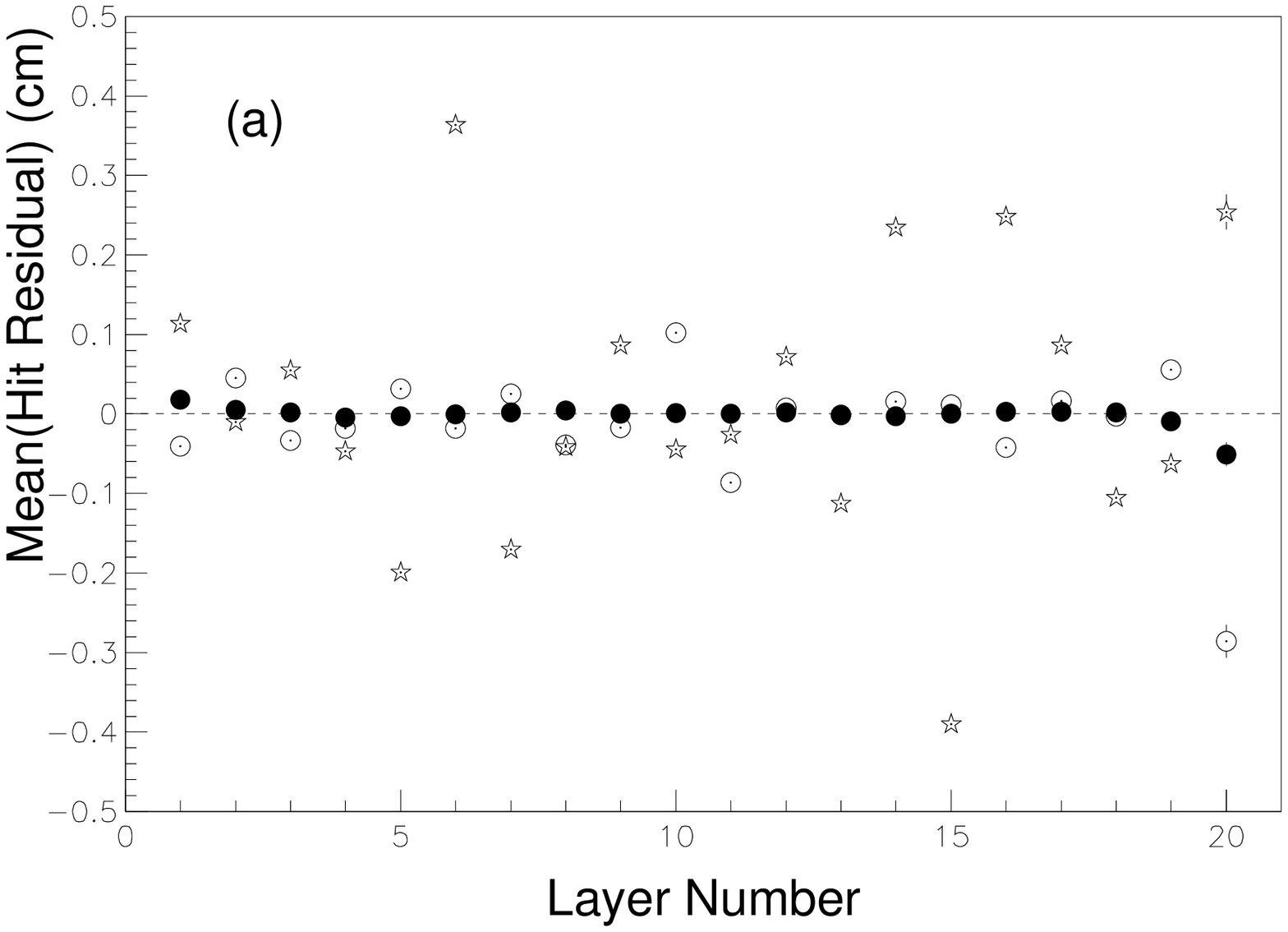,width=14cm,height=8.5cm}}
      {\psfig{figure=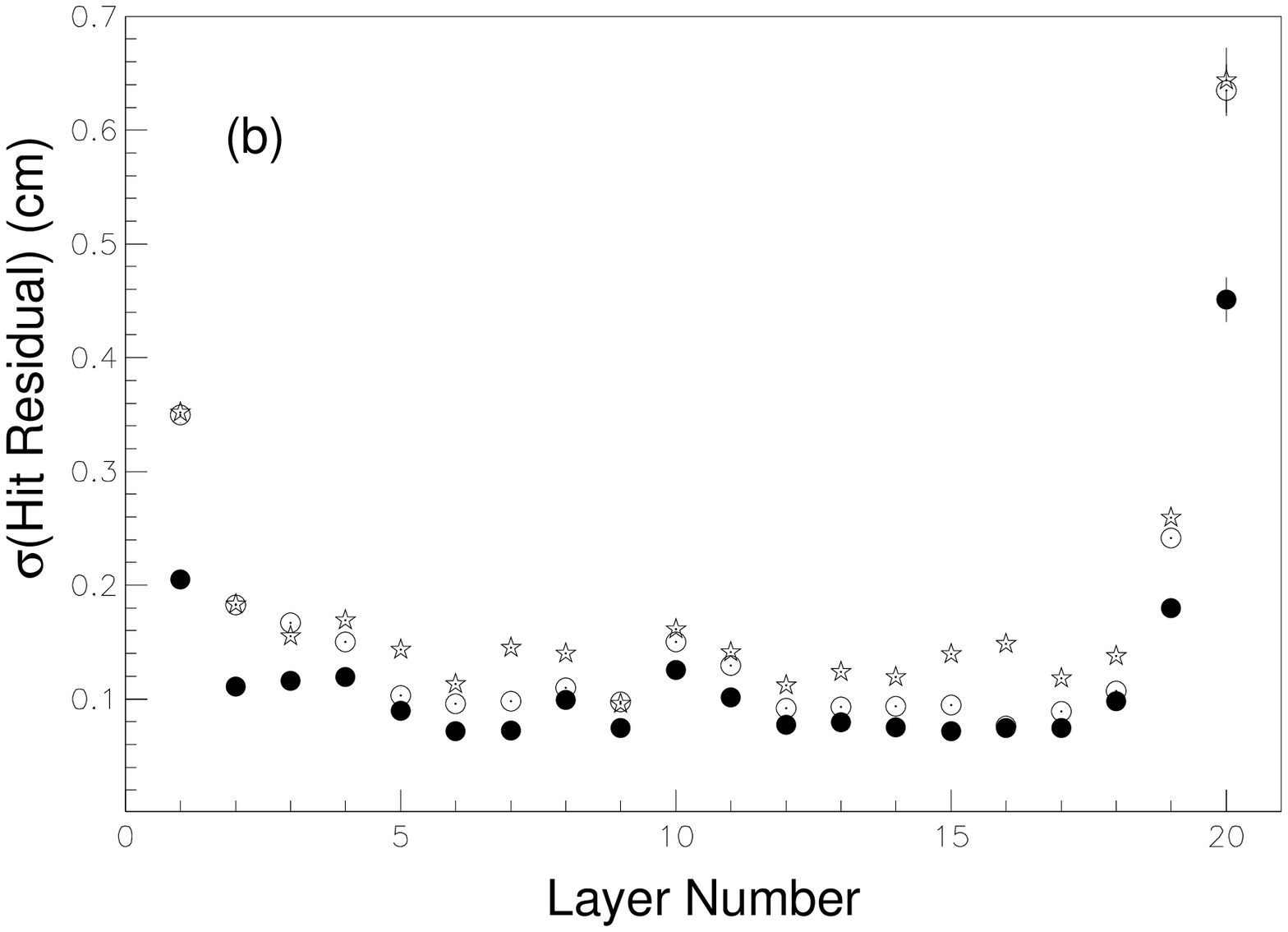,width=14cm,height=8.5cm}}
}
\caption{(a) Means of
hit residual and (b)
standard deviations as a function of layers in $y-z$ plane.
Points are for the mechanical survey alone (open star), the alignment 
using the global fit (open circle), 
and the alignment using the Kalman filtering (solid circle).}
\label{alicomp}
\end{center}
\end{figure}

After the Kalman filtering and the curvature correction,
the accuracy of alignment 
becomes better than 50$\mu$m, estimated from the mean residual shift. 
Fig.~\ref{alicomp} shows distributions 
of the mean and standard deviation of the
hit residuals for
the mechanical survey alone, 
the alignment using the global fit, and the alignment using
the Kalman filtering fit. 
%
%

The mean of hit residual after 
the mechanical survey alone shows
a variation as large as 1cm. After the alignment using the
global fit, the variation becomes smaller than 400$\mu$m.
After the alignment using the Kalman filtering fit, 
the variation becomes at most 50$\mu$m.
The standard deviations of the hit residual also exhibit such
improvements of fiber alignment as the mean values.
There are a few layers with large standard deviations of hit residual,
compared with other layers.
These are the outer most layers and indeed bent more than the rest
of inner layers. 

\begin{figure}[h]
\begin{center}
\mbox{\psfig{figure=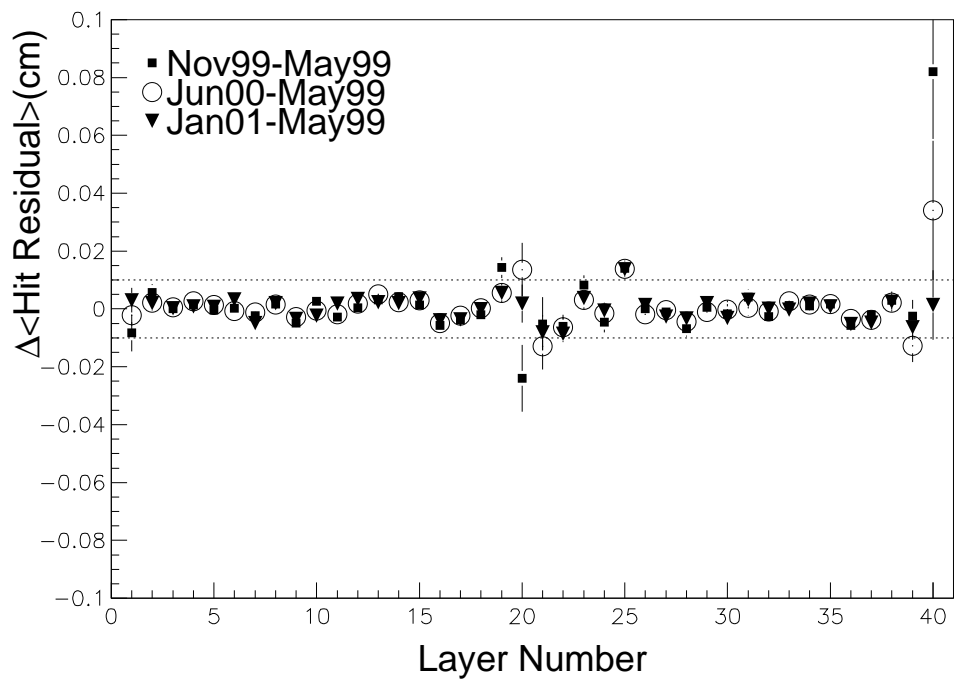,width=14cm}}
\caption{The variation in the mean of hit residual, 
obtained from cosmic-ray data, 
as a function of layers in $x-z$ plane (1-20) and $y-z$ plane (21-40). 
The variation is taken relative to May 1999.
The dotted line represents the variations of $\pm$100$\mu$m.}
\label{restab}
\end{center}
\end{figure}

The alignment is checked using periodically taken cosmic-ray data.
Fig.~\ref{restab} shows the variation in the mean values of hit residuals 
since May 1999, and the variation is found to be less than 100$\mu$m. 
%
%

\begin{figure}[h]
\begin{center}
\mbox{\psfig{figure=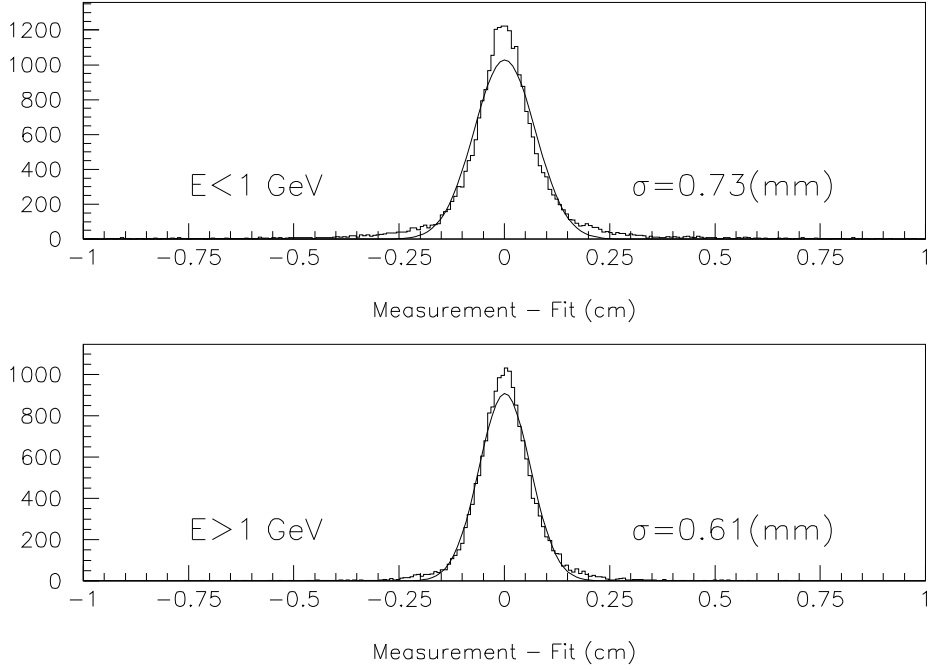,width=14cm}}
\caption{The hit residual 
distributions using the neutrino beam data
with the Kalman filtering fit. The curve is a single Gaussian fit to the data
with $\sigma$ $=$ 0.73mm (0.61mm) for muon energy E$_\mu$ $<$ 1GeV ( E$_\mu$ $>$ 1GeV).}
\label{kalres}
\end{center}
\end{figure}

\begin{figure}[h]
\begin{center}
\mbox{\psfig{figure=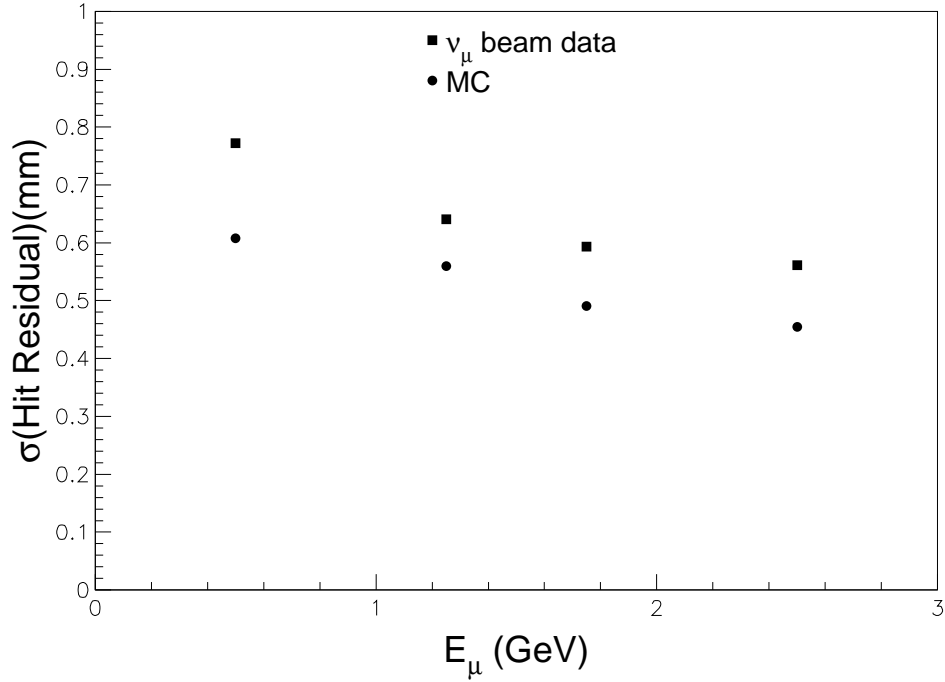,width=14cm}}
\caption{The energy dependence of position resolution.}
\label{resedep}
\end{center}
\end{figure}

The neutrino data
with the Kalman filtering fit are used to obtain the best estimation for
the position resolution.
Fig.~\ref{kalres} shows the hit residual distribution
using neutrino data.
The position resolution is 0.73mm (0.61mm) for muon energy E$_\mu$ $<$ 1GeV ( E$_\mu$ $>$ 1GeV). The overall position resolution is 0.64$\pm$0.07mm.
%
%
Fig.~\ref{resedep} shows the energy dependence of the resolution.
The effect of multiple scattering on the position resolution decreases 
as the muon energy increases. The MC prediction explains the 
overall energy dependence. The difference between data and
MC, $\sim$100$\mu$mm,  is regarded as the systematic error of our
detector. The position resolution of 0.6mm is regarded as that
of the SciFi without the multiple scattering effect
%
%

 \subsection{Track Finding Efficiency}

The efficiency for finding a single muon track is 
estimated from cosmic-ray and MC events. 
The cosmic-ray events are selected by requiring
associated activities in both upstream and downstream VETO counters.
Furthermore, tracks are required to have at least 15 SciFi
hit layers from the upstream to the downstream.
In order to study
dependence of the track finding efficiency on track length,
we artificially eliminate hits in the upstream side of the track.
For example, if it is desired to estimate 
the efficiency of the track with ten SciFi hit layers using 19-layer long
track, 
the first to the nineth hit layers are eliminated
before applying a track finding algorithm.
For MC events, 
a muon track is required to have at least three hit-layers 
within the SciFi fiducial volume, and tested by a track finding algorithm.

\begin{figure}[h]
\begin{center}
\vspace{-0.5in}
\mbox{\psfig{figure=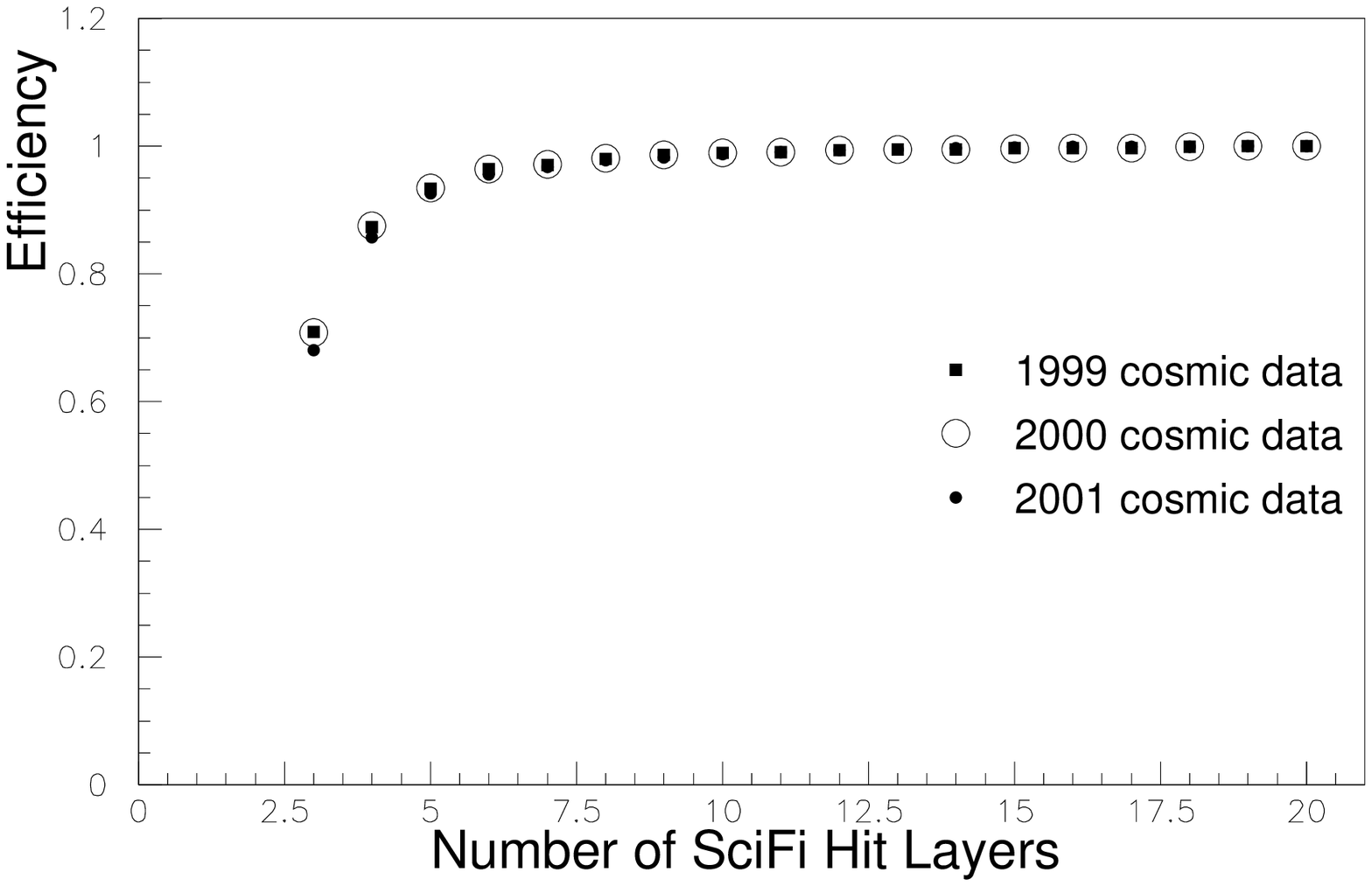,width=14cm}}
\caption{The primary track finding efficiency
as a function of track length using cosmic ray data. Solid rectangle, open circle,
and solid circle are for 1999, 2000, and 2001 cosmic ray data, respectively.
The track length is given in units of number of SciFi layers.
}
\label{costreff}
\end{center}
\end{figure}

Fig.~\ref{costreff} shows  a comparison of single track finding efficiency 
as a function of the track length
in unit of number of SciFi layers for the cosmic ray data.
%
%
Track finding inefficiencies 
come from failures in the hit finding and 
imperfection of the track reconstruction.
The track finding efficiency 
is $\sim$70\% ($\sim$87\%) for the track length of three (four) 
hit SciFi layers 
and becomes close to 100\% for longer tracks.

An average track finding efficiency for the cosmic ray events is estimated by
weighting the track length distribution obtained from the MC.
We find the average track finding efficiency to be 93.6$\pm$0.8\%
for cosmic-ray
tracks with three hits or more.
The average track finding efficiency is estimated
to be 92.7$\pm$0.5\% for the MC events.

\subsection{Vertex Reconstruction}

An accurate fiducial volume is crucial for measuring the cross 
section of neutrino in the SciFi. Therefore, the event vertex has to
be determined precisely.

\begin{figure}[h]
\begin{center}
\mbox{\psfig{figure=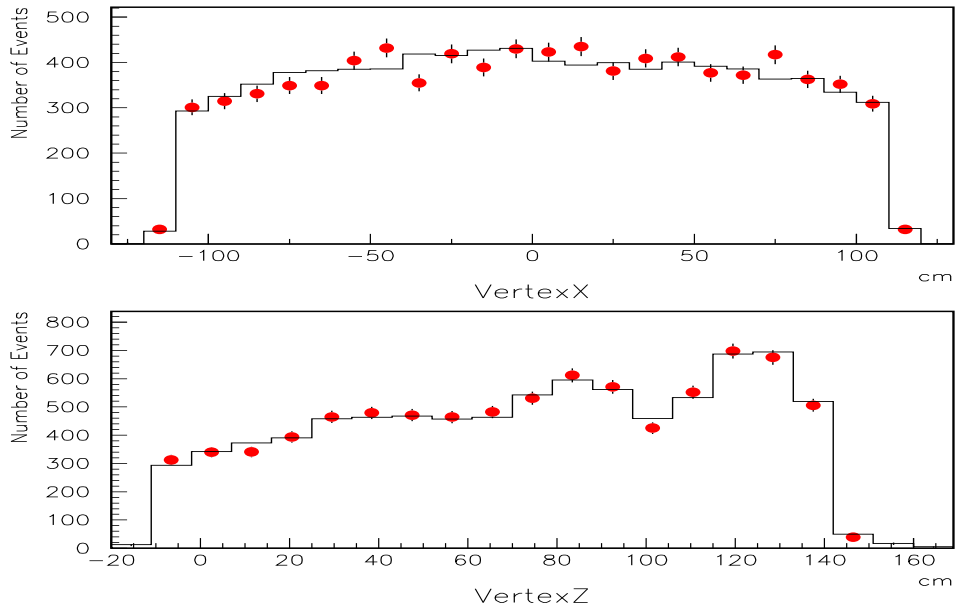,width=14cm,height=9cm}}
\caption{A comparison of $x$- and $z$-vertex distributions
between data and MC for both single and multiple track events.}
\label{vtxcomp}
\end{center}
\end{figure}

The vertex position 
is determined by the reconstructed tracks.
The event vertex for a single track event is found by simply 
taking the mid-point of the upstream water container nearest to the 
first hit in the $z$-direction of the reconstructed track.
The event vertex for a multiple track
event is determined 
by taking average of the $z$ positions at intersections of a primary
track and the rest of shorter tracks.
Fig.~\ref{vtxcomp} shows a comparison of $x$- and $z$-vertex distributions
between data and MC for both single and multiple track events.
Events with tracks ranged out at the MRD are used for the comparison
and the comparison shows a good agreement between data and MC.
%
%

\begin{figure}[h]
\begin{center}
\mbox{\psfig{figure=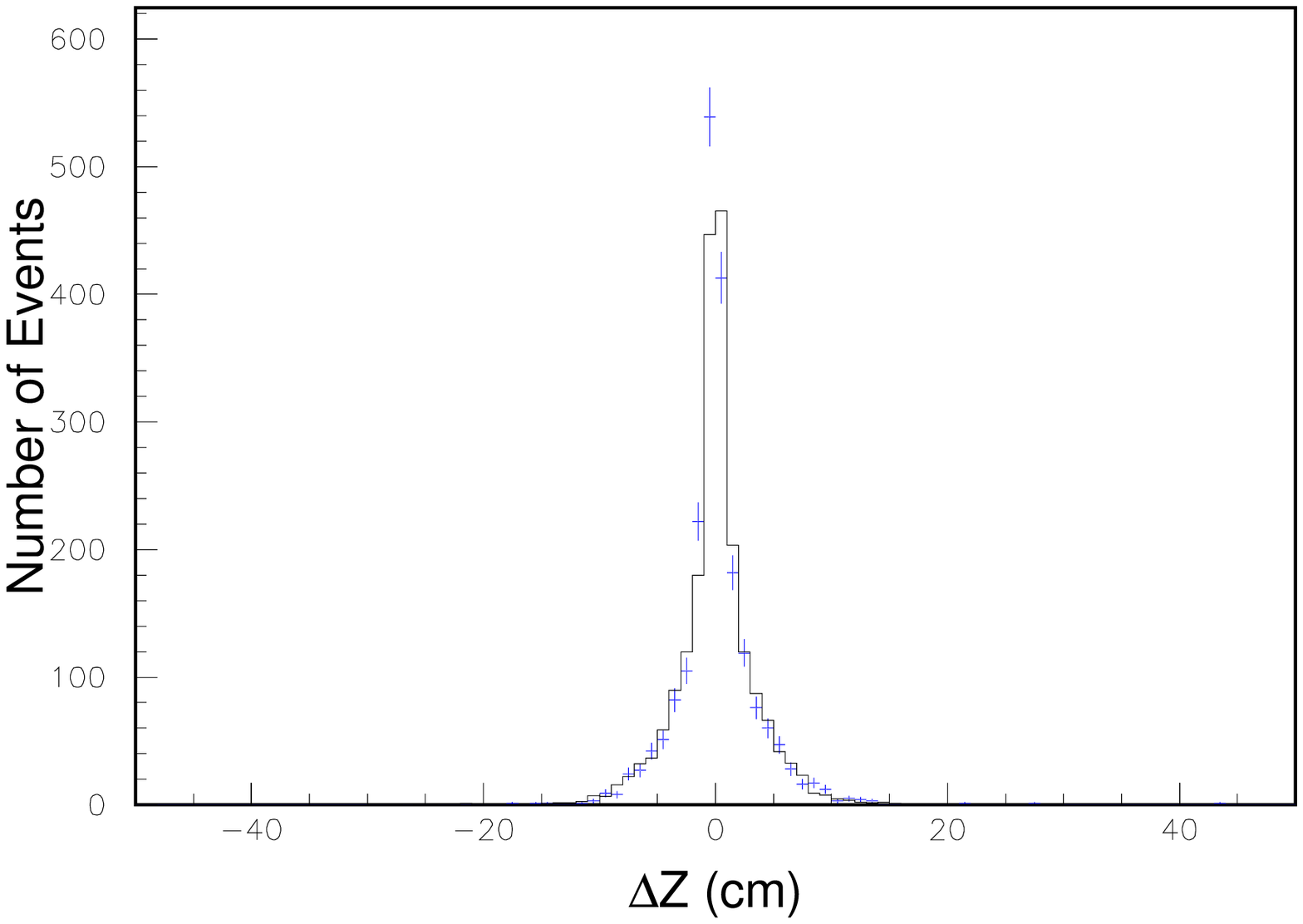,width=14cm}}
\caption{The $z$-vertex difference between vertices found in the 
$(x,z)$ and $(y,z)$ planes
for two track events. At least three hits are required for both primary
track and second track. Points are for data and the histogram is for MC.}
\label{zvertdiff}
\end{center}
\end{figure}

For two track events, $z$-vertex difference between $(x,z)$ and $(y,z)$
planes can provide a measure of goodness for the vertex reconstruction.
Fig.~\ref{zvertdiff} shows a distribution of
$z$-vertex difference between vertices found in the $(x,z)$ and $(y,z)$ 
planes for events with multiple tracks.
%
%
The event $z$-vertex is found by taking the average of the $z$ positions
determined for all such combinations of tracks. 
The event vertex in the $x$ and $y$ coordinates is 
calculated based on the event $z$-vertex
and the primary track. 

\begin{figure}[h]
\begin{center}
\vspace{-0.5in}
\mbox{\psfig{figure=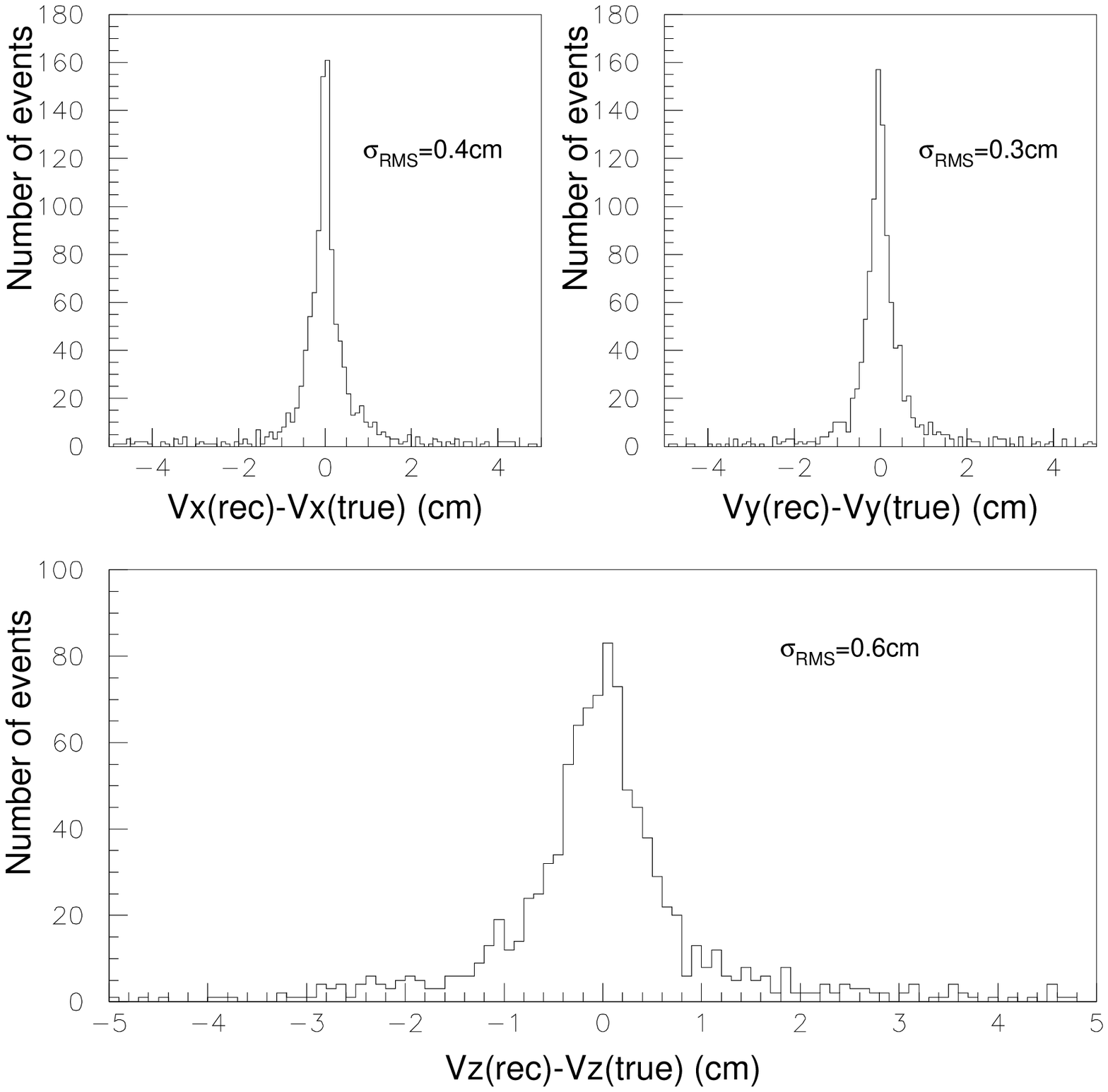,width=14cm}}
\caption{The vertex resolution of two track charged-current events.
}
\label{vtxres2trk}
\end{center}
\end{figure}

The vertex resolution is estimated by two track MC events of neutrino
charged-current interactions.
The vertex resolution is degraded mainly due to 
track mis-reconstruction. 
Fig.~\ref{vtxres2trk} shows the difference
between the measured vertex position and the true vertex position in
$x$, $y$, and $z$ for the two track events.
%
%
The estimated vertex resolution for the two track events
is found to be 
0.6 cm in $z$ and 0.4 cm in $x$ or $y$. 

\subsection{Track Counting}
An accurate reconstruction of the track multiplicity of an event is
crucial for the measurement of neutrino energy spectrum. 
Charged current quasi-elastic events have especially
simple kinematics which can be used to determine the incoming neutrino
energy.  These events always have one or two tracks depending on
whether the outgoing proton is detected.

The efficiency of finding multiple tracks is studied using the MC CCqe events,
and defined to be the fraction of reconstructed 
events relative to the generated events for a given track multiplicity.
Both the reconstructed track and the generated
track are required to have three or more hits.
The efficiency is estimated to be 93$\pm$2\% and 
86$\pm$4\% for single track CCqe events and 
two track CCqe events, respectively.
For single track events, the uncertainty in the track counting efficiency 
is dominated by the systematic error due to the gain variation of the IIT's,
and estimated to be 1\%.

For two track events, the sources of the uncertainty 
in the track counting efficiency
are difference in noise between MC and data, 
tracks due to noise hits around the vertex, 
and noise hits near the primary track.
Due to the difference in noise between data and MC, the track reconstruction
may perform differently.

In order to estimate the systematic uncertainty 
due to difference in noise contamination between the data and MC, 
noise contamination is varied in the MC by 
$\pm$50\% from the default value. 
After the variation, track finding is performed again. 
We find the track counting efficiency varies by at most 1.8\% with
respect to the default noise containmination 
in the MC for both single track and 
two track events. 

Single track events can sometimes 
be reconstructed as two track events because of noise hits
around the vertex. The size of this effect 
is studied using MC generated single track events with three
or more hits. 
The probability is calculated by counting number of reconstructed two track
events, and found to be 2.7$\pm$0.1\% 
for the MC CCqe events.

On the other hand, if there are nearby noise hits along
the primary track, single track events might be reconstructed as 
two track events.
This type of nearby noise hits contributes to the uncertainty in
track counting for two track events. 
For single track events,  
the vertex position is relocated to the middle of the track and
the track finding is repeated around the new vertex.
The uncertainty in track finding due to accidental 
noise hits around the primary track 
is found to be less than 1\%.

The track counting inefficiency can be explained by loss of tracks or 
additional fake tracks through reconstruction.
The efficiency of finding the second track 
is calculated as a function of number of hits in the SciFi layers 
using MC.
Events with two generated tracks are reconstructed, and only those with
a primary track and a vertex are selected.
The second generated track should have at least two hits.
The efficiency of finding the second track is defined to
be the ratio of number of events with two reconstructed tracks to the 
number of events with two generated tracks.

\begin{figure}[h]
\begin{center}
\vspace{-0.4in}
\vbox{
      {\psfig{figure=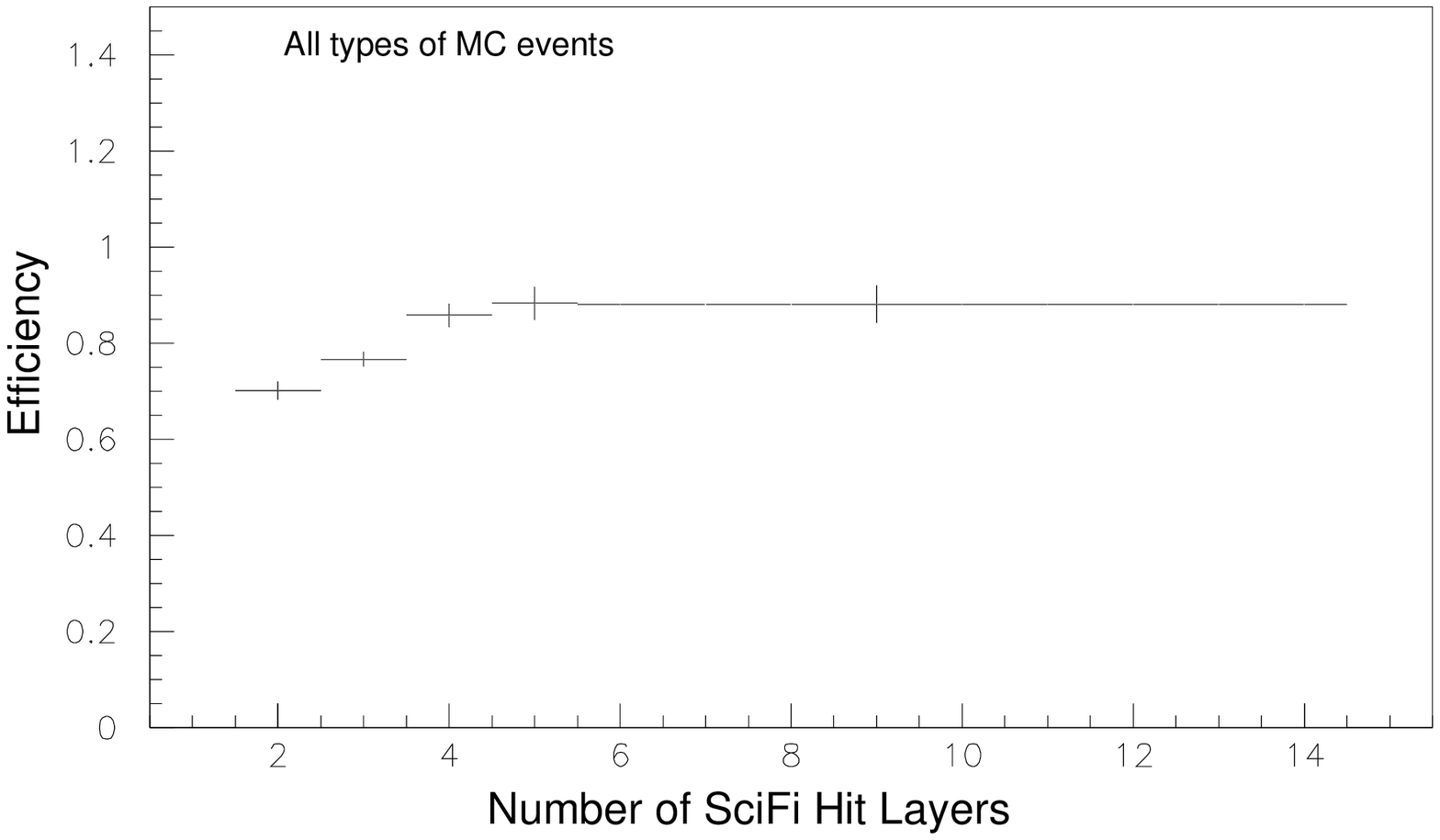,width=14cm,height=7.5cm}}
      {\psfig{figure=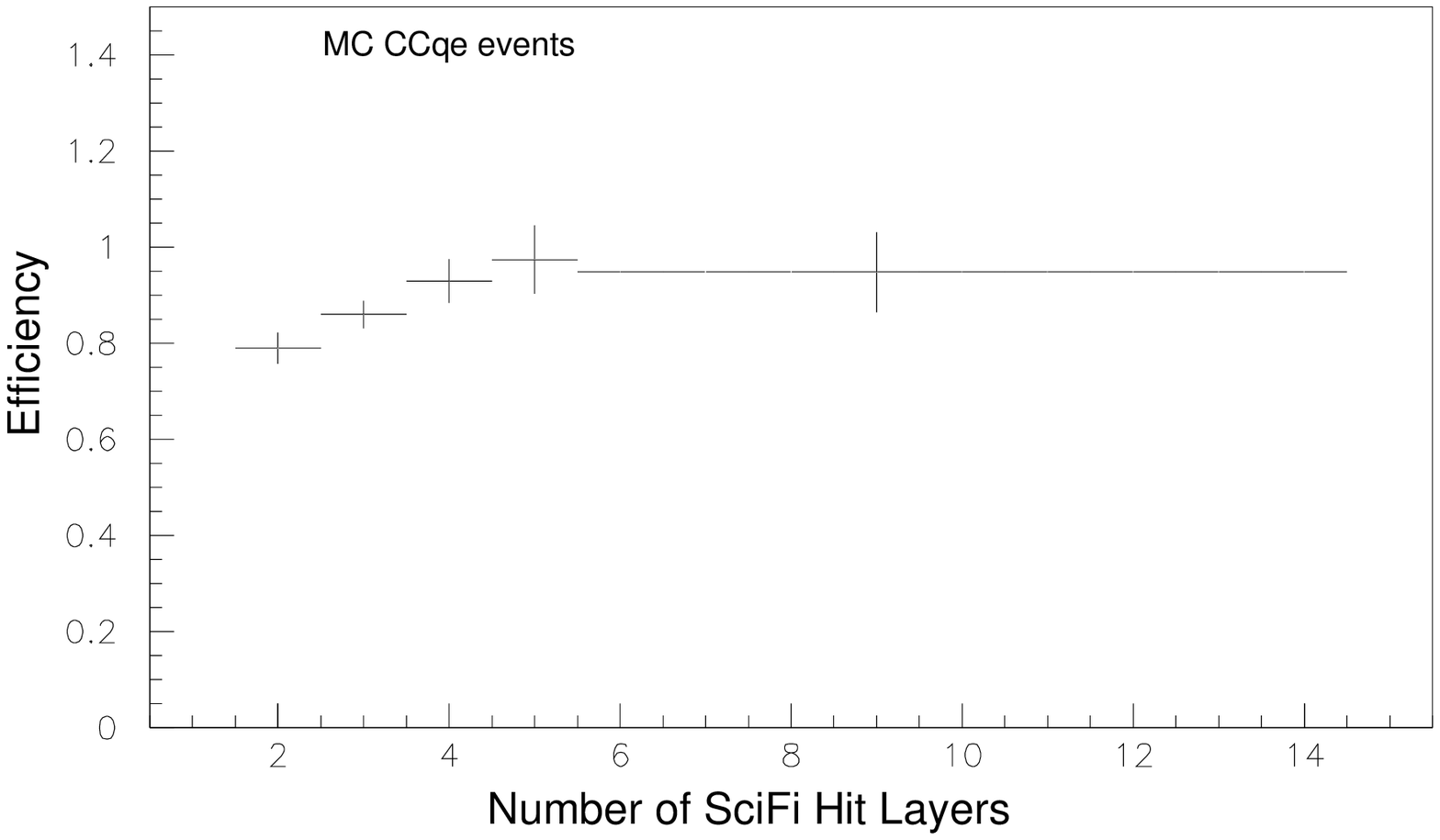,width=14cm,height=7.5cm}}
}
\caption{The efficiency of finding the second track 
as a function of number of hits in
the SciFi layers. The upper plot is for the all events and the lower
plot is for the CCqe events.}
\label{trkeff2nd}
\end{center}
\end{figure}

Fig.~\ref{trkeff2nd} shows the efficiency of finding the second track 
as a function of number of hits in the SciFi layers. The upper plot is for
all types of MC events, and the lower plot is for the CCqe events.
%
%

For the second tracks with three or more hits, the inefficiency comes from 
hit finding inefficiency and limited performance of the track 
reconstruction as for the primary track. 
For the second tracks with two hits,
the inefficiency is mainly due to 
noise hits around the vertex or near-by noise hits along the primary track.
An average efficiency of finding the second track with three or more hits 
is found to be
82.1$\pm$0.4\% for all types of MC events and 90.1$\pm$0.5\% for 
MC CCqe events.

%
%
\begin{table}[h]
\begin{tabular}{lcc} \hline\hline
Performance Quantity   & Data           & MC Events          \\ \hline
{\bf Hit finding efficiency($\epsilon_{hit}$)}   
                       &96.5$\pm$0.5\%  & 96.8$\pm$0.1\%      \\ \hline
{\bf Primary tack finding efficiency($\epsilon_{trk}$)}
                       & 93.6$\pm$0.8\% &92.7$\pm$0.5\%      \\ \hline  
{\bf Position resolution}
                       &0.64$\pm$0.07mm&0.57$\pm$0.02mm    \\ \hline   
{\bf Vertex resolution}&                &                    \\
{\bf Single track}     & N/A            & 6 mm (Transverse)  \\   
                       & N/A            & N/A                \\   
{\bf Two track}        & N/A            & 4 mm (Transverse)  \\   
                       &                & 6 mm (Longitudinal)\\ \hline  
{\bf Tracking counting}&                &                    \\    
{\bf 1-track efficiency}
                       & N/A            & 93$\pm$2\%          \\
{\bf 2-track efficiency}
                       & N/A            & 86$\pm$4\%          \\ \hline
\hline
\end{tabular}
\caption{Summary of tracking performance of the SciFi. 
Position resolution is measured using
the neutrino data. Vertex resolution is estimated using MC events.
Track counting efficiency is calculated
for the CCqe events. Errors on the hit efficiency, tracking efficiency, and
track counting efficiency are 
the quadratic sums of statistical and systematic errors.} 
\begin{center}
\label{tab:trkperf}
\end{center}
\end{table}

\section{Event Reconstruction}
\label{neutrec}
In the previous section, we described the reconstruction of tracks
and vertices. In this section, we demonstrate their performances
using distributions of typical physics variables.

\begin{figure}[h]
\begin{center}
\mbox{\psfig{figure=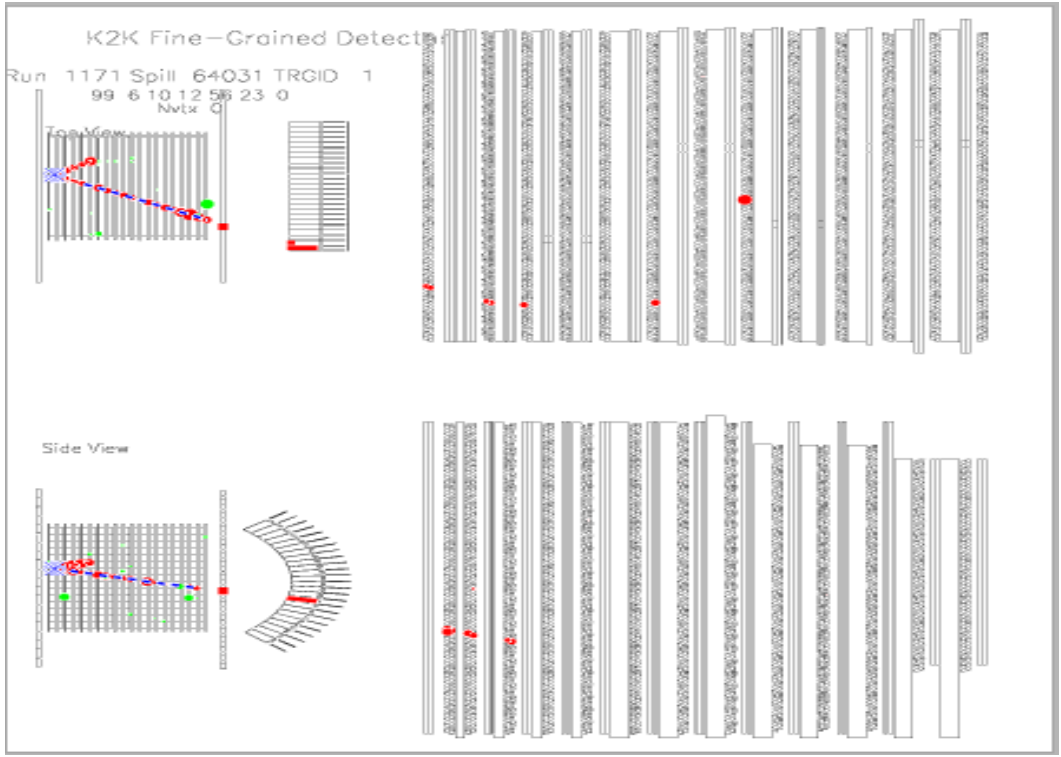,width=14cm,height=10cm}}
\caption{A candidate $\nu_\mu$ CCqe event. The long track is the muon and the
short track is most likely a proton.
}
\label{qecandi}
\end{center}
\end{figure}

Fig.~\ref{qecandi} shows a typical $\nu_\mu$ CCqe candidate event. 
The long track is the muon which leaves
hits in the VETO and LG, and ranges out at the MRD. 
The short track is most likely a proton. 
%
%

\begin{figure}[h]
\begin{center}
\vspace{-0.5in}
\mbox{\psfig{figure=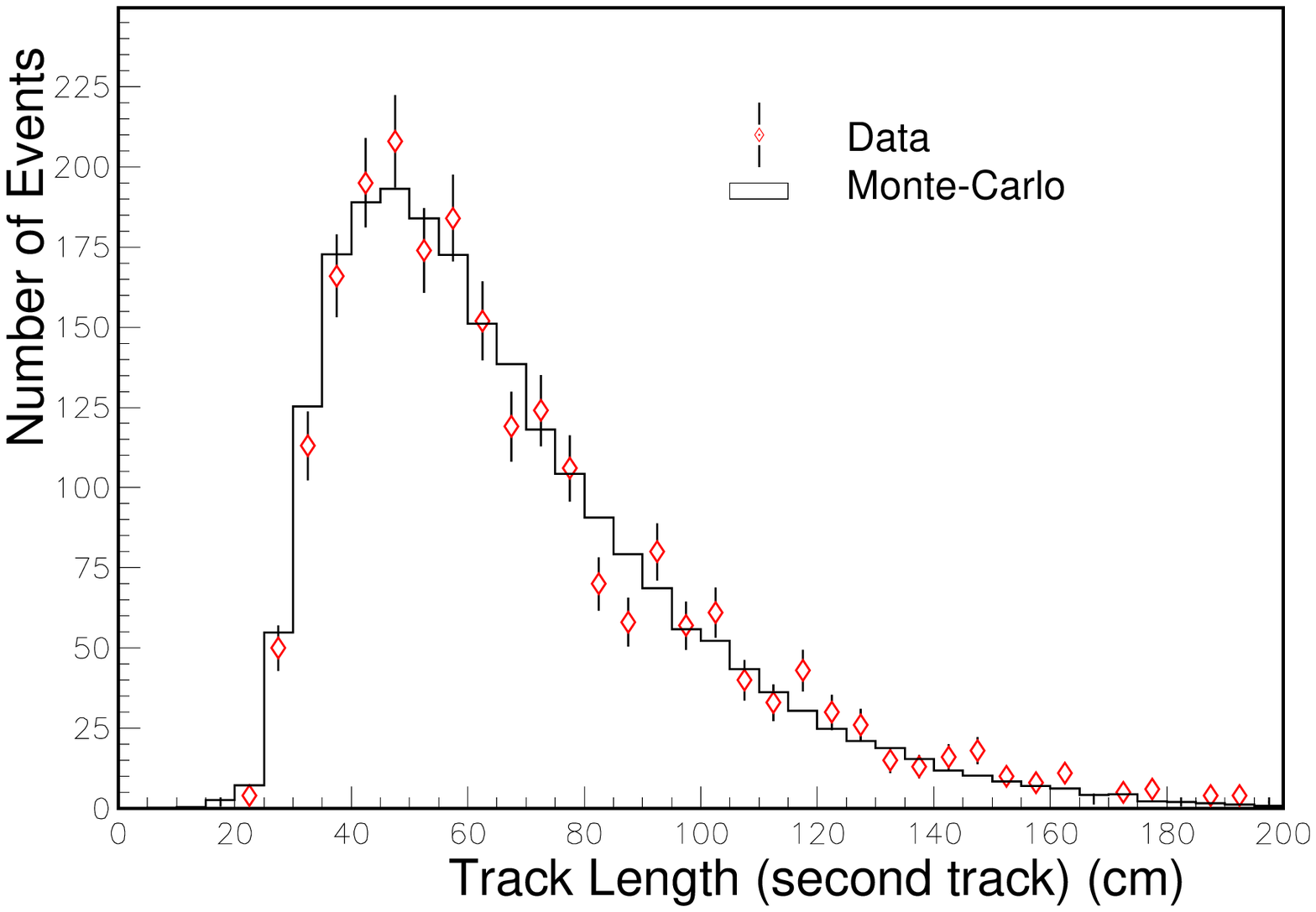,width=14cm}}
\caption{A comparion of track length for the proton between neutrino
data and MC events in two track sample.} 
\label{pleng}
\end{center}
\end{figure}

Most of the protons from neutrino interactions have relatively low momenta and 
thus make short second-tracks in the SciFi.
Fig.~\ref{pleng} shows the track length distribution for the second 
tracks
in the 2-track sample. 
Both Monte Carlo prediction and data agree well, demonstrating the validity of
the neutrino event generation and the tracking method. 
%
%

\section{Conclusions}
\label{concls}
The K2K long-baseline neutrino experiment has taken data for three years. 
The SciFi is used to identify charged particle tracks 
in the K2K near detector.

The SciFi hits are calibrated using cosmic ray data and electro-luminescent
calibration data. MC simulation reproduces the distribution of 
the number of hit pixels
in the cosmic ray data very well. The stability of the SciFi is
checked periodically by obtaining the distribution of hit efficiency and 
calculating the mean of the hit pixel distribution.
The stability is better than 5\% and 1.2\%
in the mean of pixel distribution and the hit efficiency, respectively.

A track reconstruction algorithm, optimized for charged
current neutrino interactions, is employed to find tracks and event
vertices. The track finding efficiency is monitored and 
found to be stable throughout the data-taking period. 
Using the estimated vertex resolution, 
the vertex reconstruction alogirthm is found to be good enough for
defining the event fiducial volume with an accuracy better than
1\%.
A track fit using a Kalman filtering technique is used for 
the best estimation of hit resolution, and 
found to improve the fiber
alignment.

Accurate track counting is crucial for the measurement of neutrino
energy spectrum and in achieving the physics goals of 
the K2K experiment. 
The track counting uncertainty is found to be at 2\% and 4\% level for the
single track and two track events, respectively.
Overall tracking performance of the SciFi is summarized 
in Table.~\ref{tab:trkperf}.

The performance of the SciFi tracker is well demonstrated by physics 
variables of neutrino interaction.
The distributions of variables for muon track and proton track
show a good agreement between data and MC events.

\ack{
The vital contributions of the KEK staff and the technical staffs of the
participating institutions are gratefully acknowledged. 
This work is supported
by the Ministry of Education, Science
and Culture of Japan, 
the U.S. Department of Energy, the National Science Foundation of the U.S.,
the Korea Research Foundation, the Korea Science and Engineering Foundation,
 and the CHEP in Korea.
}

\bibliographystyle{unsrt}


\end{document}